# PAC study of the static and dynamic aspects of metal atom inside $C_{60}$ cage

S.K. Das[1*], R. Guin[1], D. Banerjee[1], P. Das[2], T. Butz[3], V. S. Amaral[4], J.G. Correia[5], M. B. Barbosa[5]

[1] Accelerator Chemistry Section (Bhabha Atomic Research Centre), Variable Energy Cyclotron Centre, 1/AF Bidhannagar, Kolkata 700064, India
[2] Variable Energy Cyclotron Centre, Kolkata 700064, India
[3] Fakultät fürPhysik und Geowissenschaften, Universität Leipzig, Linnéstr.5, 04103 Leipzig, Germany
[4] Departamento de Física e CICECO, Universidade de Aveiro, 3810-193 Aveiro, Portugal
[5] ITN, Sacavém, Portugal and ISOLDE-CERN

*Author for correspondence; email:satyen@vecc.gov.in

## Abstract

30 keV $^{111m}$Cd and 50 keV $^{199m}$Hg beams from ISOLDE were used to implant on preformed target of $C_{60}$ with the thickness of 1 mg/cm$^2$. Endofullerene compounds, viz., $^{111m}$Cd@$C_{60}$ and $^{199m}$Hg@$C_{60}$ formed during implantation were separated by filtration through micropore filter paper followed by solvent extraction. Dried samples of the endofullerene compounds were counted for the coincidence for the 151-245 keV cascade of $^{111m}$Cd and 374-158 keV cascade for $^{199m}$Hg for the TDPAC measurement on a six LaBr$_3$ detector system coupled with digital electronics. The results indicate a single site occupied by the Cd atom in the fullerene cage with fast relaxation component which means that the Cd atom does move inside the cage at room temperature. The quadrupole interaction frequency and asymmetry parameter of the Cd atom occupying the site in $C_{60}$ are $\omega_Q$=8.14(42) Mrad/s and $\eta$= 0.42(9) respectively with a site population of 28%. Fast relaxation constant is 0.003 ns$^{-1}$ with a site contribution of 72%. On the other hand, Hg atom has two static sites of quadrupole frequencies, $\omega_{Q1}$=281.60(16.9) Mrad/s and $\omega_{Q2}$=202.30(22.7) Mrad/s. Respective asymmetry parameters are $\eta_1$=0.1(1) and $\eta_2$=0.2(1). The static components have a contribution of 40% with the rest 60% coming from the fast relaxing component.

## Introduction

After the discovery of fullerene[1], the compounds in this group which received much attention in basic research and in applications are the endofullerenes. The inner diameter of the $C_{60}$ cage is ~4Å which is large enough to accommodate any atom or ion across the periodic table. The behaviour of an atom or a cluster of atoms inside the fullerene cage makes such compounds suitable for various applications in the area of superconductivity, lasers, and ferroelectric materials[2]. A very promising application of these compounds lies in its medical use[3] where the trapped radioactive atom inside the inert carbon cage is not in direct contact with the biological system – a prerequisite for a nuclear medicine. An optimistic conjecture has been made to use endofullerenes for the storage of nuclear waste by entrapping the radioactive atoms inside the cage[4, 5]. The self-healing property of C-C bonds is helpful in preserving the cage even if the C-C bonds are cleaved by the radiation from radioactive atoms inside the cage. Besides the applications mentioned above, basic research interest is enormous as far as the chemical bonding of the atom inside the cage is concerned. This is the most important aspect in the study of endofullerenes.

The most well-known fullerene, $C_{60}$, has three degenerate lowest unoccupied molecular orbitals (LUMOs), energetically slightly above the highest occupied molecular orbitals (HOMOs)[6]. These can accommodate up to six electrons and hence, metal ions with valency up to six can be accommodated inside the cage. However, a trivalent metal ion contributing three electrons to make LUMOs half-filled stabilizes the endofullerene. Since the inner side of the cage is electrostatically slightly positively charged as the π-electrons of the $sp^2$-carbon atoms (in fullerene carbon hybridisation is $sp^{2.27}$) are bulged outside, a negatively charged ion is expected to be stabilized inside the cage. The energy consideration due to the transfer of electrons between the entrapped atom and the carbon atoms of the cage is the decisive factor in the stabilization of the endofullerenes.

Endofullerenes are synthesised by two different ways: (i) simultaneous formation of fullerene in the arc vaporisation or laser ablation from the metal doped graphite electrode, (ii) posterior insertion of the desired atom by implantation on the pre-existing fullerene. In the process (i), thermodynamics decides the formation of the endfullerene. However, in the case of the post-insertion using ion implantation, the atom once enters the cage has no chance to come out. So it is only the barrier that decides the formation of the endofullerene. Once the atom enters the cage, the question remains how it behaves inside the cage. An EPR study[7] indicates that while a nitrogen atom inside the $C_{60}$ cage does not interact with the carbon atoms and remains in the central position of the cage, Cu in Cu@$C_{60}$ occupies a well defined (minimum potential) position inside the cage. This indicates that there is a bond of the Cu atom with C atoms. An NMR study [8] and the MEM/Rietveld method[9] for imaging of diffraction data indicates that the metal atoms inside the cage rotate. A small rotational barrier in such cases can help to control the rotation by temperature. This property of the endofullerenes can be used to develop interesting molecular devices. It is thus important to know the states of the atom inside the fullerene cage in terms of its position in the cage, charge states, dynamics etc. These parameters would be of great help to understand the mechanism of endofullerene formation which in turn would tell the route for the synthesis of the required endofullerene in significant quantities for the applications mentioned above.

TDPAC, a nuclear probe technique, relies on the interaction of the nuclear quadrupole moment with the electric field gradient (EFG) produced by the surrounding electrons. The angular correlation between γ-rays emitted in a cascade is exploited to obtain information on the electron distribution through the measurement of the EFG. The EFG and its asymmetry are the fingerprints of the electronic state of the probe atom. One thus can identify the position of the atom inside the fullerene cage with the support of theoretical calculations using e.g. density functional theory. In case there is rotation or grazing movement of the probe inside the cage, it will show up as a time dependent perturbation of the angular correlation.

With the motivation to understand the behaviour of the atom inside the fullerene cage we carried out the experiment starting from the synthesis of the endofullerene $^{111m}Cd@C_{60}$ and $^{199m}Hg@C_{60}$ by using the $^{111m}Cd$ and $^{199m}Hg$ beams from ISOLDE, CERN on the preformed $C_{60}$ target. After separation of the endofullerene from other products containing the $^{111m}Cd$ and $^{199m}Hg$ activity using a suitable chemical method described below, TDPAC measurements were carried out for the separated endofullerenes in order to investigate the static and dynamic behaviour of the atom inside the fullerene cage. Although Cd and Hg fall within the same group, Hg atom, being the bottom-most element, is expected to exhibit a different behaviour from Cd-atom because of the Inert-Pair effect. This chemical inertness of the Hg-atom and its effect on the dynamic aspect of this atom inside the fullerene cage has also been looked into with the aid of the present hyperfine tool TDPAC.

**Theory:**
The PAC method relies on the hyperfine interaction of nuclear magnetic dipole or electric quadrupole moments with extra nuclear magnetic fields or Electric Field Gradient (EFG). In case of electric quadrupole perturbation of the γ-γ angular correlation, the coincidence measurements generate the quadrupole frequency ($\omega_Q$) along with its distribution (δ) and asymmetry parameter (η) defined by $\eta = \left|\frac{V_{xx}-V_{yy}}{V_{zz}}\right|$. The TDPAC measurements were performed using the 151-245keV γ-γ cascade of the $^{111m}$Cd nucleus after the IT decay of the $^{111m}$Cd parent and 374-158 keV γ-γ cascade of the $^{199m}$Hg nucleus after the IT decay of the $^{199m}$Hg parent. The relevant decay scheme of these two nuclei is shown in Fig. 1. The perturbation of the angular correlation function results from the interaction between the electric quadrupole moment intermediate nuclear level of the probe nuclei (245 keV level with τ=85ns and *I=5/2* for $^{111m}$Cd and 158 keV with τ=2.45ns and *I=5/2* for $^{199m}$Hg) and the electric field gradient (EFG) around the probe atom.

The Nuclear Quadrupole Interaction (NQI) of the I = $5/2$ intermediate state leads to a splitting with Eigen values[10]

$$E_1 = -2r\cos\left(\frac{\phi}{3}\right)$$

$$E_2 = r\cos\left(\frac{\phi}{3}\right) - \sqrt{3}\,r\sin\left(\frac{\phi}{3}\right)$$

$$E_3 = r\cos\left(\frac{\phi}{3}\right) + \sqrt{3}\,r\sin\left(\frac{\phi}{3}\right)$$

With

$\cos\phi = q/r^3$

$r = \text{sign}(q)\sqrt{|p|}$

$p = -28\left(1 + \eta^2/3\right)$

$q = -80(1 - \eta^2)$

The three precession frequencies are:

$$\omega_1 = (E_2 - E_3)\omega_Q$$

$$\omega_2 = (E_1 - E_2)\omega_Q$$

$$\omega_3 = (E_1 - E_3)\omega_Q$$

Where $E_1$, $E_2$ and $E_3$ are the three eigenvalues (in units of $\hbar * \omega_Q$) corresponding to the splitting of the intermediate state (I=5/2) of the probe due to the interaction of quadrupole moment (Q) of the intermediate state with the electric field gradient tensor ($V_{zz}$) with the quadrupole frequency as:

$$\omega_Q = \frac{eQV_{zz}}{40\hbar}$$

The perturbation function is thus:

$$G_2(t) = a_0 + a_1\cos\omega_1 t + a_2\cos\omega_2 t + a_3\cos\omega_3 t$$

The experimental data were fitted with this function modified using finite distributions of $\omega_Q$. In the present case we have used Lorentzian distributions. The final form of the theoretical function in which the experimental data were fitted is given by:

$$G_2(t) = a_0 + \sum_{n=1}^{3} a_n \exp(-\omega_n \delta t) \times \exp(-1/2\, \omega_n^2 \tau^2) \cos(\omega_n t) \tag{1}$$

The exponential damping terms attribute to the finite resolving time characterized by a Gaussian distribution with standard deviation $\tau$ and the Lorentzian frequency distribution with relative width parameter $\delta$. The coefficients $a_n$ have dependence on the nuclear radiation parameters and the asymmetry parameter.

Probes occupying more than one site (say, '$i$' sites) gives rise to a total perturbation function with different amplitudes $a_i$

$$G_2(t) = \sum_{i=1}^{n} a_i G_2^i(t)$$

When the electric field fluctuates in magnitude and direction, it gives rise to time-dependant EFG since the axis of symmetry undergoes constant change. This fluctuating electric field is realized by a parameter called the correlation time ($t_c$) which signifies the time separating two appreciable interactions between the nuclear quadrupole moment and the EFG. In case of isotropic fluctuations fast enough to fulfil both conditions $t_c \ll \tau$ and $t_c \ll \frac{2\pi}{\omega_0}$, the Abragam-Pound theory[11] leads to an exponential attenuation factor:

$$G_k(t) = \exp(-\lambda_k t) \tag{2}$$

Where the relaxation constant $\lambda_k$ is described by the following equation:

$$\lambda_k = \frac{3}{80}\left(\frac{eQ}{\hbar}\right)^2 \langle V_{zz}^2 \rangle t_c k(k+1) \left[\frac{4I(I+1) - k(k+1)}{I^2(2I-1)^2}\right]$$

For the intermediate I=5/2 state, the relaxation constant (in addition to $\lambda_0 = 0$) takes the following form:

$$\lambda_2 = 0.063 \left(\frac{eQ}{\hbar}\right)^2 \langle V_{zz}^2 \rangle t_c = 100.8 \langle \omega_Q^2 \rangle t_c$$

The correlation time reflects the dynamic behaviour of the atom inside the fullerene cage. The smaller correlation time signifies a faster movement of the atom inside the cage and vice versa. This aspect has been investigated in our present study.

**Experimental**

$C_{60}$ targets were prepared on 10x8 mm$^2$ whatman-42 filter paper by soaking the toluene solution of $C_{60}$ on the filter paper mentioned above and subsequently drying the filter paper. The target amount was approximately 1 mg/cm$^2$. 30 keV beam of $^{111m}$Cd and 50 keV beam of $^{199m}$Hg available from ISOLDE were implanted on the $C_{60}$ target to prepare $^{111m}$Cd@$C_{60}$ and $^{199m}$Hg@$C_{60}$ endofullerene in two separate experiments. Beam current was ~ $10^9$ ions/s. Total fluxes of ions collected was $3\times10^{12}$ ions for different samples. Six irradiations were carried out for each ion. The irradiated fullerene target was treated with toluene to dissolve out the fullerene from the filter paper. Any decomposed products or agglomerate of fullerene formed due to the irradiation of fullerene with energetic ions were removed by micron filter paper (pore size is 0.5 micron). This filter paper contained a large amount of radioactivity and only a small fraction of the order of 1% was available in the filtered solution. This filtered solution was treated for the solvent extraction with 6M HCl solution to remove the $^{111m}$Cd or $^{199m}$Hg attached to fullerene exohedrally or the ions attached outside the $C_{60}$ cage. The atoms attached to the exo products are expected to dissolve in 6M HCl medium[12]. However the atoms in trapped inside the cage are shielded by the inert carbon atoms and are unaffected by the acid and thus the endofullerene products remain in the organic phase. The toluene fraction thus contains only the endofullerenes along with pure fullerene molecules. Toluene was evaporated to get the dried endofullerene fraction which was then counted on the TDPAC setup[13] consisting of six LaBr$_3$(Ce) detectors coupled to the digital electronics. $^{111m}$Cd ($t_{1/2}$=49m) and $^{199m}$Hg($t_{1/2}$=43m) have similar half lives. Each sample was counted for three hours nearly four half lives of the parent nucleus.

**Results and Discussion**

It was found that the activity of either $^{111m}$Cd or $^{199m}$Hg in the chemically separated fraction of endofullerene was two orders of magnitude less than the irradiated sample before chemical separation. In other words, the yield of endofullerene is less than a percent. Most of the radioactivity was found in the filter paper used for the separation of the decomposed products of fullerene suspended in toluene. Energetic $^{111m}$Cd or $^{199m}$Hg ions when interact with fullerene loses its energy and when it reaches few tens of eV (barrier for entrance), the ions open up the hexagon ring and get trapped inside the cage. During the process of energy loss there is damage of fullerene molecules which may agglomerate to bigger size and suspend in toluene medium. During this process, a lot of Cd and Hg ions get trapped in the agglomerates of fullerene molecules. These are separated by the micropore filter paper. So the toluene fraction after solvent extraction contains only the endofullerene part along with pure fullerene molecules. Fig. 2 shows the TDPAC spectrum for the $^{111m}$Cd@$C_{60}$ endofullerene and Fig. 3 shows that for the $^{199m}$Hg@$C_{60}$. The data are fitted with the static interaction as mentioned in the equation (1) and a time dependent part mentioned in eqn (2). Fitting has been performed using the least square fitting code WINFIT. It is seen that the TDPAC spectrum for $^{111m}$Cd@$C_{60}$ could be fitted with a static and one time-dependent interaction. On the other hand, the data for Hg could be fitted with two static interactions and one time-dependent interaction. Table -1 show the TDPAC results obtained by fitting the spectra for both probes. Results indicate that Cd atoms occupy a definite site with 30% population and the rest 70% is due to fast relaxation caused by the fluctuating field around the probe nucleus. The origin of this fluctuating is expected to arise from the movement of the probe atoms inside the cage. In case of Hg, the data fitting with minimum $\chi^2$ indicates the presence of two definite static sites where the probe $^{199m}$Hg can reside. Total population of these static states is 40% with 16% of the state with $\omega_Q$=281.6Mrad/s and 24% with $\omega_Q$=202.3 Mrad/s. The rest 60% is for the fluctuating field. The observation of two static sites is in contrast with the Cd ion, a member of the same group of Hg. However, the difference in chemical behaviour between these two elements is that Cd exits in only one valence state - II and Hg can be in valence states of both I and II. Therefore the existence of two sites for Hg could be conjectured due to the existence of Hg in two valence states which give rise to two interactions. Now the question is how both static and dynamic interaction can exist at the same time for these probes. It would be difficult to assume that atoms inside the cage can give both static and dynamic interactions. The only reason could be that when Cd and Hg atoms, with some kinetic energy, interact with $C_{60}$, there is damage in the $C_{60}$ molecules[14] and during this process the metal atoms may be trapped in a cluster of damaged $C_{60}$ molecules. The bigger agglomerates are separated by the micropore filter paper which contains almost 99% of the total activity of the probes. However a fraction of these clusters which are smaller in dimension might not be separated by the filter papers and remained suspended in toluene even after filtration. These small clusters entrapping the probe atoms give rise to the static fraction of the hyperfine interaction. Again the two static parts for Hg-atom are explained by its two different valence states present in the fullerene clusters. The dynamic fraction is obviously due to the movement of the metal atoms inside the cage of $C_{60}$. This is possible when the potential on the inner surface of $C_{60}$ is relatively flat indicating no preferred place where Cd or Hg can attach. Similar situation has been observed in case of $La_2@C_{80}$[15]. Since $\lambda_2 \propto \langle\omega_Q^2\rangle t_c$, it follows that $t_c$ for Hg is 1-2 orders of magnitude smaller than that for Cd. A comparison of the rate of relaxation between Cd and Hg shows a faster movement of Hg than Cd despite Hg being heavier than Cd. The reason is that Hg having inert pair effect, the s- electrons of Hg have much less affinity to interact with the carbon atoms of $C_{60}$. This makes Hg more loosely bound and thus has faster motion as indicated by the higher relaxation constant λ. Theoretical calculation in the line of mapping the potential of the metal atom on the inner surface of $C_{60}$ is in progress to corroborate the experimental results for the endofullerene. To check the nature of products formed by the energetic ions with $C_{60}$, a typical sample of the filter paper along with the damaged products of fullerene was counted. This showed very different TDPAC results. This is shown in Fig. 4. Wide frequency distribution indicates that the products do not have the definite chemical nature. They are no way expected to be endofullerenes.

## Conclusion

Energetic beam of nuclear probes viz. [111m]Cd and [199m]Hg were used to implant on the preformed $C_{60}$ target to produce the respective endofullerenes [111m]Cd@$C_{60}$ and [199m]Hg@$C_{60}$. They were separated from other products by chemical method. TDPAC of the endofullerene components indicate that in both the cases there exists a time dependent component of the hyperfine interaction. This shows that both Cd and Hg atoms move inside the fullerene cage but with different speed. Faster motion of the Hg atom compared to that of Cd has the origin that Hg has inert pair effect causing weaker interaction of Hg with the carbon atoms than that of Cd atom. Static part of the TDPAC spectrum possibly arises because of some specific decomposed products of fullerene attached with the probe nuclei. Measurement at different temperature will be interesting to see the change in the motion of the atoms inside the fullerene cage.

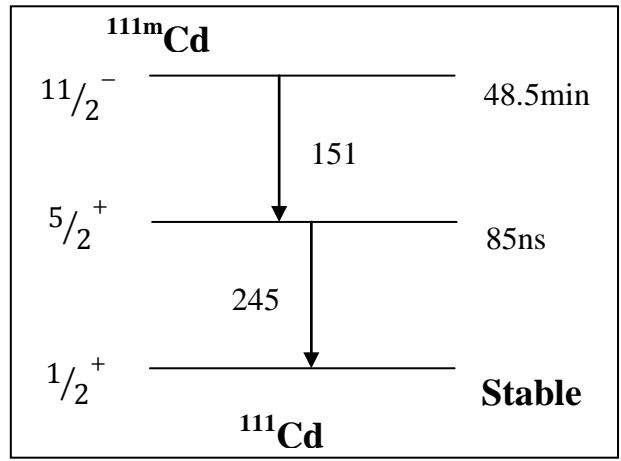

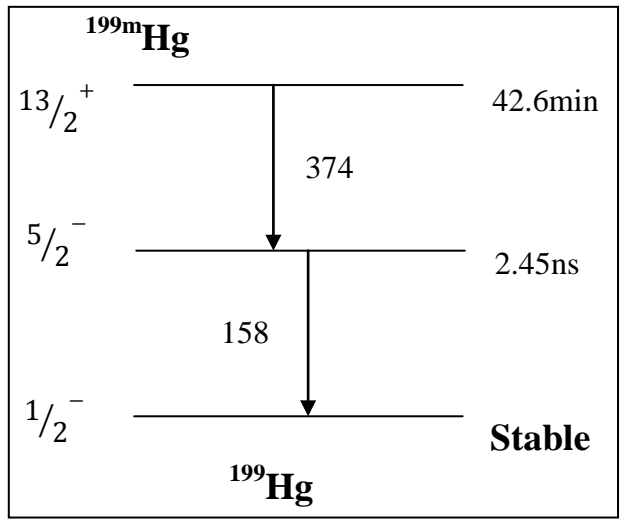

Figure 1: (a) Decay scheme of $^{111m}$Cd and (b) Decay scheme of $^{199m}$Hg

| Sample | $\omega_Q$ (Mrad/s) | H | δ (%) | Fast λ (ns$^{-1}$) |
|---|---|---|---|---|
| $^{111m}$Cd | 8.14(42) | 0.42(9) | 9.3(5.7) | 0.003(3) |
| $^{199m}$Hg | 281.6(16.9) | 0.1(1) | 0.01(2) | 0.06(9) |
| | 202.3(22.7) | 0.2(1) | 0.08(5) | |

Table 1. TDPAC results for the endofullerene $^{111m}$Cd@C$_{60}$ and the decomposed products on the filter papers

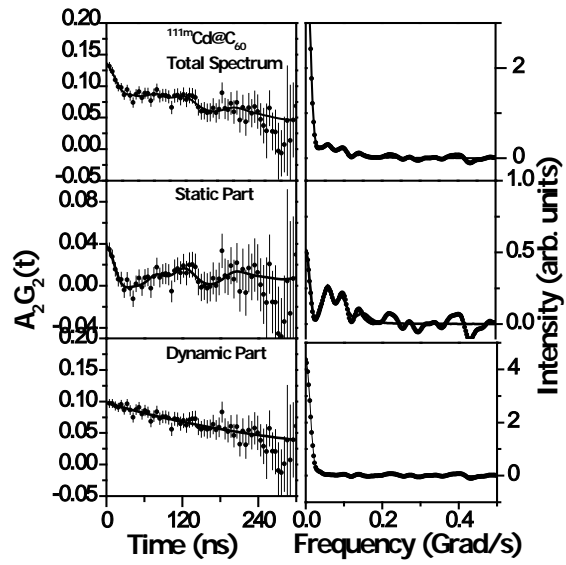

Figure 2. TDPAC spectrum for the $^{111m}$Cd@C$_{60}$ endofullerene (left) and the corresponding Fourier spectrum (right).

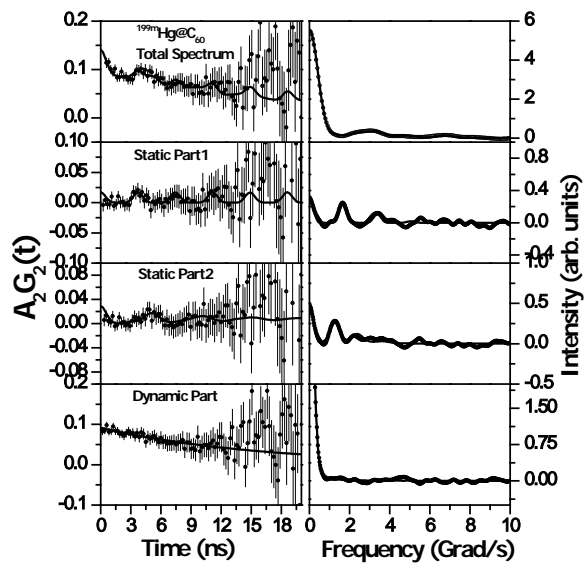

Figure 3. TDPAC spectrum for the C$_{60}$ decomposed products on the filter paper (left) and the corresponding Fourier spectrum (right).

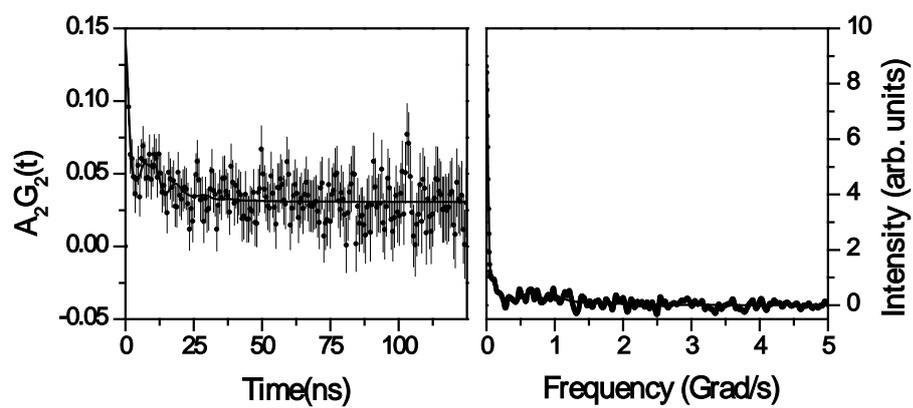

Figure 4. TDPAC spectrum for the $C_{60}$ decomposed products on the filter paper (left) and the corresponding Fourier spectrum (right).